# Rare recombination events generate sequence diversity among balancer chromosomes in *Drosophila melanogaster*

Danny E. Miller[a,b], Kevin R. Cook[c], Nazanin Yeganehkazemi[a], Clarissa B. Smith[a], Alexandria J. Cockrell[a], R. Scott Hawley[a,b,1] and Casey M. Bergman[d,1]

[a] Stowers Institute for Medical Research, Kansas City, MO 64110, USA
[b] Department of Molecular and Integrative Physiology, University of Kansas Medical Center, Kansas City, Kansas 66160, USA
[c] Department of Biology, Indiana University, Bloomington, IN 47405, USA
[d] Faculty of Life Sciences, University of Manchester, Manchester M13 9PT, United Kingdom

[1]To whom correspondence should be addressed:
R. Scott Hawley
Stowers Institute for Medical Research
1000 E 50th Street
Kansas City, MO, 6411
Email: rsh@stowers.org

Casey M. Bergman
Michael Smith Building
University of Manchester
Oxford Road
Manchester, M13 9PT
United Kingdom
Email: casey.bergman@manchester.ac.uk

**Short title:** Structure and evolution of a balancer chromosome (<50 characters)

**ABSTRACT**

Multiply inverted balancer chromosomes that suppress exchange with their homologs are an essential part of the genetic toolkit in *Drosophila melanogaster*. Despite their widespread use, the organization of balancer chromosomes has not been characterized at the molecular level, and the degree of sequence variation among copies of any given balancer chromosome is unknown. To map inversion breakpoints and study potential sequence diversity in the descendants of a structurally identical balancer chromosome, we sequenced a panel of laboratory stocks containing the most widely used X-chromosome balancer, *First Multiple 7* (*FM7*). We mapped the locations of *FM7* breakpoints to precise euchromatic coordinates and identified the flanking sequence of breakpoints in heterochromatic regions. Analysis of SNP variation revealed megabase-scale blocks of sequence divergence among currently used *FM7* stocks. We present evidence that this divergence arose by rare double crossover events that replaced a female-sterile allele of the *singed* gene ($sn^{X2}$) on *FM7c* with wild type sequence from balanced chromosomes, and propose that many *FM7c* chromosomes in the Bloomington *Drosophila* Stock Center have lost $sn^{X2}$ by this mechanism. Finally, we characterize the original allele of the *Bar* gene ($B^1$) that is carried on *FM7* and validate the hypothesis that the origin and subsequent reversion of the $B^1$ duplication is mediated by unequal exchange. Our results reject a simple non-recombining, clonal mode for the laboratory evolution of balancer chromosomes and have implications for how balancer chromosomes should be used in the design and interpretation of genetic experiments in *Drosophila*.

**SIGNIFICANCE STATEMENT**

Balancer chromosomes are highly rearranged chromosomes that suppress recombination and are an important tool in *Drosophila* genetics. Balancers were synthesized in the pre-genomic era, and thus both their precise molecular structures and the level of diversity among balancer chromosomes are unknown. Here, we characterize the breakpoints of inversions on the *X*-chromosome balancer *FM7*, and provide evidence that rare double crossover events with balanced homologs can lead to substantial sequence diversity among balancer chromosomes in use today. We also provide genomic evidence that unequal exchange between duplicated regions underlies reversion at the *Bar* locus on *X*-chromosome balancers. Our work demonstrates the power of genome sequencing to understand the molecular nature of classical genetic resources and phenomena.

**INTRODUCTION**

Balancer chromosomes are genetically engineered chromosomes that suppress crossing over with their homologs and are used for many purposes in genetics, including construction of complex genotypes, maintenance of stocks, and estimation of mutation rates. Balancers typically carry multiple inversions to suppress genetic exchange, recessive lethal or sterile mutations to prevent their propagation as homozygotes, and dominant mutations as markers for their easy identification. First developed for use in *Drosophila melanogaster*, balancer chromosomes remain one of the most powerful tools for genetic analysis in this species (Ashburner *et al.* 2005).

Despite their widespread use, very little is known about the organization of *Drosophila* balancer chromosomes at the molecular level. Since their original syntheses decades ago, balancers have undergone many manipulations including the addition or removal of genetic markers. Additionally, rare recombination events can cause spontaneous loss of deleterious alleles on chromosomes kept over balancers in stock, as well as loss of marker alleles on balancer chromosomes themselves (Ashburner *et al.* 2005). Likewise, recent evidence has shown that sequence variants can be exchanged between balancer chromosomes and their wild type homologs *via* gene conversion during stock construction or maintenance (Cooper *et al.* 2008; Blumenstiel *et al.* 2009). Thus, substantial variation may exist among structurally identical balancer chromosomes due to various types of sequence exchange.

To gain insight into the structure and evolution of balancer chromosomes, we have undertaken a genomic analysis of the most commonly used *X*-chromosome balancer in *D. melanogaster, First Multiple 7* (*FM7*). We have focused on *FM7* because this *X*-chromosome balancer series lacks lethal mutations and can therefore easily be sequenced in a hemizygous or homozygous state. In addition, the *FM7* chromosome has been shown to pair normally along most of its axis with a standard *X*-chromosome, providing a structural basis for possible exchange events (Gong *et al.* 2005). Moreover, although details of how early balancers in *D. melanogaster* were created are not fully recorded, the synthesis and cytology of the *FM7* series is reasonably well documented (Ashburner *et al.* 2005).

The earliest chromosome in the *FM7* series, *FM7a*, was constructed using two progenitor *X*-chromosome balancers, *FM1* and *FM6*, to create a chromosome carrying three inversions – *In(1)sc$^8$*, *In(1)dl-49*, and *In(1)FM6* – relative to the wild type configuration (Merriam 1968; 1969) (**Figure 1A**). Subsequently, a female-sterile allele of *singed* (*sn$^{X2}$*) was introduced onto *FM7a* to create *FM7c*, which prevents the loss of balanced chromosomes carrying recessive lethal or female-sterile mutations (Merriam and Duffy 1972). More recently, versions of *FM7a* and *FM7c* have been generated that carry transgene insertions that allow balancer genotypes to be determined in embryonic or pupal stages (Casso *et al.* 2000; Le *et al.* 2006; Abreu-Blanco *et al.* 2011; Lattao *et al.* 2011; Pina and Pignoni 2012).

To identify the inversion breakpoints in *FM7* balancers and to study patterns of sequence variation that may have arisen since the origin of the *FM7* series, we sequenced genomes of eight *D. melanogaster* stocks carrying the *FM7* chromosome (four *FM7a* and four *FM7c*). We discovered several megabase-scale regions where *FM7c* chromosomes differ from one another, which presumably arose *via* double crossover (DCO) events from balanced chromosomes. These DCOs eliminate the female-sterile $sn^{X2}$ allele in the centrally located *In(1)dl-49* inversion and are expected to confer a fitness advantage by allowing propagation of *FM7* as homozygotes in females. We show that loss of the $sn^{X2}$ allele is common in *FM7c* chromosomes by screening other *FM7c*-carrying stocks at the Bloomington *Drosophila* Stock Center. We also identified the breakpoints of the $B^1$ duplication carried on *FM7* and provide direct molecular evidence for the role of unequal exchange in the origin and reversion of the $B^1$ allele (Sturtevant and Morgan 1923; Sturtevant 1925; Muller 1936; Peterson and Laughnan 1963; Gabay and Laughnan 1973). Our results provide clear evidence that the common assumption that balancers are fully non-recombining chromosomes is incorrect on a historical timescale and that substantial sequence variation exists among balancer chromosomes in circulation today.

RESULTS

**Identification of *FM7* inversion breakpoints**

The inversions carried by *FM7* that confer the ability to suppress recombination were generated by *X*-ray mutagenesis and characterized using genetic and cytogenetic data in the pre-genomic era, and thus the precise locations and molecular nature of their breakpoints remain unknown. To better understand the genomic organization of *FM7* chromosomes, we used whole-genome sequencing to identify breakpoints for the three inversions present on *FM7*: *In(1)sc$^8$*, *In(1)dl-49*, and *In(1)FM6* (**Figure 1A**). Based on cytological data, it is known that both breakpoints of *In(1)dl-49* lie in euchromatic regions (Painter 1934; Hoover 1938; Lindsley and Zimm 1992). However, for both *In(1)sc$^8$* and *In(1)FM6*, one breakpoint is euchromatic and the other lies in centric heterochromatin (Sidorov 1931; Patterson 1933; Muller and Prokofyeva 1934; Patterson and Stone 1935; Grell and Lewis 1956; Lindsley and Zimm 1992).

Our general strategy to identify breakpoint regions is as follows. We sequenced eight *FM7*-carrying stocks to approximately 50-fold coverage with paired-end Illumina data and mapped reads to the *D. melanogaster* reference genome (see summary statistics in **Supplemental Table 1**). We identified clusters of split or discordantly mapped reads from all stocks in the vicinity of expected breakpoint locations based on previous cytological data, then performed *de novo* assembly of split/discordant reads and their mate-pairs (reads from the other end of the same paired-end sequenced fragments). Breakpoint contigs identified by sequence analysis were then used to design PCR amplicons that span breakpoints, and resulting PCR amplicons were Sanger sequenced to verify breakpoint assemblies. Using this approach, we were able to map euchromatic breakpoints of all three inversions on the *FM7* chromosome to reference genome coordinates, as well as characterize the sequence composition of the heterochromatic breakpoints for both *In(1)sc$^8$* and *In(1)FM6* (**Figure 1B**).

The distal breakpoint of the X-ray-induced *In(1)sc$^8$* inversion has been localized near bands 1B2-3 between the *achaete (ac)* and *scute (sc)* genes (Sidorov 1931; Patterson 1933; Muller and Prokofyeva 1934; Patterson and Stone 1935; Campuzano et al. 1985; Lindsley and Zimm 1992). We identified a cluster of split/discordant reads in *FM7* stocks around X:276,500 (predicted band 1A7) of the type expected in the vicinity of an inversion breakpoint. Split/discordant reads from ±1.5 kb around the putative *In(1)sc$^8$* inversion breakpoint (which map to the A and B regions) and their mate-pairs (which map to the C and D regions) were extracted from all *FM7* strains, pooled together and assembled to identify candidate A/C and B/D breakpoint sequences. BLAST analysis of the resulting assembly revealed two contigs of 506 bp and 551 bp. The euchromatic components of these contigs mapped to nucleotides X:276,417–276,422 in the Release 5 genome sequence between *ac* and *sc*, within an intron of *CG32816*. The heterochromatic components of these contigs contained copies of the 1.688 satellite DNA repeat (Hsieh and Brutlag 1979) that covers approximately half of the *X*-chromosome centric heterochromatin (Lohe *et al.* 1993). The locations and sequences of candidate breakpoints for *In(1)sc$^8$* were used to design PCR primers that yielded amplicons in all stocks carrying *In(1)sc⁸* but not in stocks lacking this inversion (**Supplemental Table 2**). Sanger sequencing of PCR amplicons spanning breakpoint regions confirmed the sequence of A/C and B/D *de novo* assemblies. Comparison of A/C and B/D fragments revealed a 6-bp sequence (TTTCGT) from the *ac–sc* region that is present at both breakpoint junctions, suggesting the X-ray-induced inversion event created a small, staggered break at the euchromatic end. Our candidate A/C and B/D breakpoint regions also had strong BLAST hits to an *In(1)sc⁸* A/C junction from the *Dp(1;f)1187* mini-chromosome and the corresponding wild type A/B junction identified in a previous study (Glaser and Spradling 1994). Both our A/C fragment and that obtained by Glaser & Spradling (1994) map the euchromatic part of the distal *In(1)sc⁸* breakpoint to the same location in the *D. melanogaster* euchromatin and contain 1.688 satellite DNA in their heterochromatic part.

*In(1)dl-49* is an X-ray-induced inversion (Muller 1926) with both distal and proximal breakpoints in euchromatic regions at bands 4D7–E1 and 11F2–4, respectively (Painter 1934; Hoover 1938; Lindsley and Zimm 1992). We identified clusters of split/discordant reads for the distal breakpoint near X:4,791,300 (predicted band 4D5) and for the proximal breakpoint from approximately X:13,321,200–13,321,900 (predicted band 11F6). These candidate breakpoint intervals were also identified using Breakdancer (Chen *et al.* 2009), an independent method which is able to predict inversions that have two euchromatic breaks. We extracted split/discordant reads within ±1.5 kb of each of the putative *In(1)dl-49* breakpoint intervals plus their mate-pairs, pooled reads from both breakpoints, then performed *de novo* assembly followed by PCR and Sanger sequencing. As expected, PCR amplification was successful in stocks carrying *In(1)dl-49* but failed in stocks lacking *In(1)dl-49* (**Supplemental Table 2**). Sanger sequencing verified the sequence of the A/C and B/D breakpoint assemblies. Both the proximal and distal breakpoints were found in unique genomic regions, with the distal break occurring between X:4,791,293–4,791,295 in an intron of *CG42594* and the proximal break occurring from X:13,320,887–13,321,245 in an intergenic region between *SET domain containing 2 (Set2)* and *Neuropilin and tolloid-like (Neto)* (**Figure 1B**). The breakpoint in the A/C fragment contained a small 3-bp duplication that is not present in the reference genome, suggesting repair of a small staggered break during the inversion process. A 358-bp deletion was found in the B/D fragment, possibly due to resection during the repair event, which explains why the split/discordant reads for the proximal breakpoint mapped to an interval in the reference genome rather than to a single point.

The distal euchromatic breakpoint of the X-ray-induced *In(1)FM6* was reported to be near bands 15D–E (Grell and Lewis 1956; Lindsley and Zimm 1992). We identified a cluster of split/discordant reads near X:16,919,300 (predicted band 15D3) in *FM7* stocks and used these reads and the corresponding reads from the other end of the same paired-end sequenced fragments for *de novo* assembly. PCR using primers based on the two resulting putative A/C and B/D contigs validated that this breakpoint was present in all *FM7* stocks but not in stocks that lack the *In(1)FM6* inversion (**Supplemental Table 2**), and Sanger sequencing of amplicons verified the predicted breakpoint sequences.

Euchromatic components of the A/C and B/D fragments map to the same location within an intron of *CG45002* and reveal that the inversion breakpoint introduced a 1-bp deletion (X:16,919,304) **(Figure 1B)**. The heterochromatic part of the A/C fragment contains sequence from the transposable element *HMS-Beagle* (Snyder *et al.* 1982), and the heterochromatic part of the B/D fragment contains 18S rDNA sequence, consistent with the proximal breakpoint being in *X*-chromosome centric heterochromatin (Tartof and Dawid 1976). The fact that the heterochromatic regions in the A/C and B/D fragments are not the same sequence suggests either a complex breakage/repair event following irradiation or post-inversion rearrangement of sequences at either the A/C or B/D breakpoint. Nevertheless, the structure of the euchromatic junctions for the *In(1)sc$^8$*, *In(1)dl-49*, and *In(1)FM6* inversions carried on *FM7* show that X-ray-induced mutagenesis can often generate rearrangements with relatively precise breakpoints.

**Recombination generates sequence variation among *FM7* chromosomes**

It is widely believed that balancers seldom undergo recombination (Theurkauf and Hawley 1992; Hughes *et al.* 2009), giving rise to the idea that they should diverge from each other clonally and thus accumulate deleterious mutations under Muller's Ratchet (Araye and Sawamura 2013). However, previous studies have shown that sequence exchange can occur, albeit rarely, both into and out of balancer chromosomes (Cooper *et al.* 2008; Blumenstiel *et al.* 2009), although the frequency and genomic scale of such events is unknown. To test if ongoing sequence exchange between balancers and homologous chromosomes has occurred since the original synthesis of the first *FM7* chromosome, we identified variants present on only one of the eight *FM7* chromosome in our sample. Unique variants that differentiate one *FM7* from all others in our sample can arise by either by *de novo* mutation or by recombination events that donate sequence from homologous chromosomes to balancers (by either gene conversion or crossing over). However, crossing over is the only mechanism that can explain the large contiguous tracts of sequence variation that are unique to individual *FM7* chromosomes.

As shown in **Figure 2B**, we observe megabase-scale tracts of unique variation on three of the eight *FM7* chromosomes (*FM7c-5193*, *FM7c-36337*, *FM7a-23229*), superimposed on a relatively even distribution of unique variants along the remainder of the chromosome. Notably, all of these tracts of unique variation are contained within the *In(1)dl-49* inversion and span the *sn* locus, and are found only in *sn$^+$* stocks. These tracts of variants were not caused by placement of the *sn$^{X2}$* allele onto *FM7a* to create *FM7c* (Merriam and Duffy 1972), since *FM7c*'s marked with *sn$^{X2}$* (*FM7c-616*, *FM7c-3378*) do not differ substantially in their SNP profile from *FM7a*'s in the *sn* region (**Supplemental Figure 1B**). In fact, similarity between *FM7a* and the original *FM7c* is expected in the *sn* region since a *In(1)dl-49* chromosome was a progenitor of *FM7a* (Merriam 1968; 1969), the *sn$^{X2}$* allele arose on a *In(1)dl-49* chromosome (Bender 1960), and a *sn$^{X2}$* marked *In(1)dl-49* was used as the donor to move *sn$^{X2}$* onto *FM7a* to create *FM7c* (Merriam and Duffy 1972). The nature of the *sn$^{X2}$* allele was not determined in earlier studies (Paterson and O'Hare 1991), however we identified a cluster of split/discordant reads at X:7,878,402–7,878,413 that arises from the insertion of an *F*-element in the 2nd coding exon of *sn* that is present only in the *sn*⁻ stocks *FM7c-616* and *FM7c-3378*. We propose that this *F*-element insertion is the lesion that causes the $sn^{X2}$ allele. Additionally, if the tracts of variants in *FM7c-5193*, *FM7c-36337*, *FM7a-23229* arose from movement of $sn^{X2}$ onto *FM7c*, they would not be unique. Rather, they would form a haplotype shared by all other *FM7c* chromosomes, as is observed in the region surrounding the *g* locus (**Supplemental Figure 1B**). The *FM7c g* haplotype on *FM7a-23229* is unexpected, and suggests that this balancer is actually a *FM7c* that has been mislabeled as *FM7a* because of its $sn^{+}$ phenotype. Together, these results indicate that all chromosomes with large tracts of unique SNPs are *FM7c*'s that lack the $sn^{X2}$ allele.

The number of unique single nucleotide variants expected on each *FM7* chromosome if they evolved clonally and independently under *de novo* mutation alone since their origin in 1968 (Merriam 1968; 1969) to the time our lines were sequenced is approximately 150 (45 years * 26 generations/year * $22 \times 10^{6}$ bp * $5.8 \times 10^{-9}$ mutations/bp/generation (Haag-Liautard *et al.* 2007)). Shared ancestry among chromosomes in our sample, such as for the *FM7c* chromosomes that were generated several years later (Merriam and Duffy 1972), would lower the number of unique variants observed from this expectation. The number of unique variants observed for five out of eight *FM7* chromosomes (56–152 unique SNPs) is less than or nearly equal to the expected value under independent clonal evolution with *de novo* mutation alone. However, the number of unique variants observed for *FM7c-5193*, *FM7c-36337*, *FM7a-23229* (between 541–3,564 unique SNPs) is more than three times higher than expected under clonal evolution with mutation alone, suggesting that the action of additional processes such as gene conversion or crossing over is required to explain these observations. The large tracts of unique variation on *FM7c-5193*, *FM7c-36337*, *FM7a-23229* range between 1.7–3.0 Mb in length and encompass 195–356 genes. Since the average tract length of gene conversion in *D. melanogaster* is approximately 350–450 bp (Hilliker *et al.* 1994; Miller *et al.* 2012), we propose that the large tracts of unique variants on *FM7c-5193*, *FM7c-36337,* and *FM7a-23229* arose by independent DCOs from unrelated chromosomes onto different *FM7* balancer chromosome lineages that replaced $sn^{X2}$ with $sn^{+}$.

The most obvious donor for sequence exchange onto a balancer chromosome is the chromosome with which it is kept in stock. To test whether the large tracts of unique sequence variation we observe on *FM7* chromosomes are the result of recombination with their homolog in stock, we sequenced heterozygous females from the three stocks with putative DCO events (*FM7c-5193*, *FM7c-36337,* and *FM7a-23229*) and from one negative control with no putative DCO event (*FM7c-616*). If a recent exchange event occurred between the balanced chromosome and its homolog, we would expect to see a loss of heterozygosity (LOH) in the region where the two chromosomes underwent recombination. As shown in **Figure 2C**, the distribution of all SNPs (both homozygous and heterozygous variants) in heterozygous samples is high and relatively constant across the entire *X*-chromosome for three of the four stocks, with two small regions in *FM7a-23229* yielding a paucity of SNPs because of shared ancestry between all *FM7*'s and the $y^1$ chromosomes in both *ISO-1* and the balanced chromosome (see **Supplemental Figure 1C**). Analysis of heterozygous SNPs in heterozygous females (**Figure 2D**) shows a relatively uniform distribution of heterozygous SNPs across the *X*-chromosome, with clear LOH in the exact region of the predicted exchange event for *FM7c-5193*, but not for *FM7c-36337* or *FM7a-23229*. These results indicate that recent exchange between *FM7c-5193* and its balanced homolog can explain the large tract of unique variants on this chromosome. However, the predicted exchange events for *FM7c-36337* or *FM7a-23229* must have occurred sometime in the past with different chromosomes other than those with which they are currently kept in stock.

Intriguingly, all three putative DCOs are contained within the central *In(1)dl-49* inversion, occur on *FM7c* chromosomes, and replace the female-sterile $sn^{X2}$ allele that was present on the original *FM7c* (Merriam and Duffy 1972) with a wild type allele. Although DCOs fully within the *In(1)dl-49* regions are rare (Sturtevant and Beadle 1936; Novitski and Braver 1954), such events would lead to viable *FM7*-bearing gametes. Furthermore, replacement of the female-sterile $sn^{X2}$ allele with $sn^+$ would lead to derived *FM7* chromosomes with a reproductive advantage relative to the ancestral *FM7c*, and thus these rare recombination events could quickly increase in frequency in stock. To address how often loss of $sn^{X2}$ occurs in *FM7c* chromosomes, we screened and classified the *sn* phenotype in males from 630 stocks carrying a *FM7c* chromosome in the Bloomington *Drosophila* Stock Center (**Supplemental Table 3**). Of 630 stocks labeled as carrying *FM7c*, we found 82 (13%) had the revertant $sn^+$ phenotype in *B*-eyed males, consistent with loss of the female-sterile $sn^{X2}$ allele on *FM7c* chromosomes by DCO with a balanced homolog inside the *In(1)dl-49* inversion while maintained in stock.

Since at least one of the *FM7a* stocks we sequenced (*FM7a-23229*) was in reality a *FM7c* stock mislabeled as a *FM7a* stock, the lack of $sn^{X2}$ on *FM7* chromosomes could simply reflect that these chromosomes are actually *FM7a*'s that are mislabeled as *FM7c*'s, rather than true loss of $sn^{X2}$ by a DCO inside *In(1)dl-49* on *FM7c*. To resolve these alternatives, we took advantage of the fact that all *bona fide FM7c*'s are expected to carry the same allele at the *garnet* locus ($g^4$), whereas all *FM7a*'s should lack this marker. Within the mutant *g* gene on all *FM7c* (and *FM7a-23229*) chromosomes (**Supplemental Figure 1B**), we found a diagnostic 24-bp deletion that spans an intron-exon junction and results in a frame-shift in the RB and RD transcripts (FBtr0331709 and FBtr0073842), and also ablates the ATG start codon of the RF transcript (FBtr0331710). We tested 79 of the 82 revertant $sn^+$ stocks labeled as *FM7c* in Bloomington for the presence or absence of this putative $g^4$-causing deletion by PCR and Sanger sequencing. We found that 74/79 (94%) of the $sn^+$ stocks screened by PCR and sequencing carried the $g^4$ allele present on all *FM7c* chromosomes (**Supplemental Table 3**), indicating that the majority of these are *bona fide FM7c*'s and thus truly revertants. Because *g* lies outside the *In(1)dl-49* inversion and *sn* resides inside it, it is highly unlikely that one DCO event could have replaced both $sn^{X2}$ and $g^4$ in any of the five putative *FM7c* $sn^+$ stocks that lack the $g^4$ deletion. We therefore conclude that these five stocks have been mislabeled as *FM7c* when, in fact, they are actually *FM7a*'s. Thus, the vast majority of $sn^+$ stocks labeled as *FM7c* in the Bloomington *Drosophila* Stock Center are indeed *FM7c*'s, but mislabeling of *FM7* subtypes (a versus c) occurs in about of 6% of stocks. Overall, these results support the conclusion that the DCOs within the *In(1)dl-49* interval occur an appreciable frequency, endangering mutations in homologous chromosomes kept in stock over balancer chromosomes, and leading to sequence diversity among *FM7c* balancers in circulation today.

**Origin and reversion of the $B^1$ allele**

*X*-chromosome balancers including *FM7* carry the $B^1$ allele, a dominant mutation affecting eye morphology, discovered more than 100 years ago (Tice 1914). $B^1$ is an unusual allele that reverts to wild type at a high frequency in females (May 1917; Zeleny 1921) through either inter-chromosomal or intra-chromosomal unequal exchange (Sturtevant and Morgan 1923; Sturtevant 1925; Peterson and Laughnan 1963; Gabay and Laughnan 1973). $B^1$ is known to revert on *FM7* (Ashburner *et al.* 2005) and previous work suggests that $B^1$ reversion rates may be higher in inverted *X*-chromosomes (Sturtevant and Beadle 1936; Gabay and Laughnan 1973). $B^1$ has been shown to be associated with a tandem duplication of a large segment containing cytological bands 16A1–7, and $B^1$ revertants lack this duplicated segment (Muller *et al.* 1936; Bridges 1936). Muller (1936) argued that $B^1$ arose by unequal exchange between two sister chromatids or homologous chromosomes, rather than a duplicative insertion event as suggested by Bridges (1936). Muller's model for the origin of $B^1$ was supported by the work of Tsubota *et al.* (1989) who used a *P*-element-induced revertant of $B^1$ to clone the putative breakpoint of the $B^1$ duplication. These authors found a *roo* transposable element located exactly at the breakpoint between the two duplicated segments, and proposed that the $B^1$ allele originated by unequal exchange between *roo* elements located at 16A1 and 16A7, respectively, on two different homologous chromosomes (Tsubota *et al.* 1989) **(Figure 3A)**. However, the exact nature of the $B^1$ rearrangement remains to be clarified, since the 16A7 breakpoint of $B^1$ identified by Tsubota *et al.* (1989) contained a short segment of DNA not found in wild type flies. Moreover, neither the genomic extent nor gene content of the $B^1$ duplication has been investigated in the context of modern genomic data.

We identified the precise genomic limits of the $B^1$ duplication on the basis of a contiguous 203,476-bp region between X:17,228,526–17,432,002 with two-fold higher sequencing depth in all eight *FM7* stocks sequenced **(Figure 3B).** Sequences flanking the duplicated interval correspond exactly to the $B^1$ breakpoints identified by Tsubota *et al.* (1989). We found that previous uncertainty in the wild type configuration of the 16A7 $B^1$ breakpoint region reported by Tsubota *et al.* (1989) is due to inclusion of phage DNA in their sequence. The $B^1$ duplicated interval contains the *BarH1 (B-H1)* homeodomain gene that has been shown to be involved in the *Bar* eye phenotype (Kojima *et al.* 1991; Higashijima, Kojima, *et al.* 1992), plus seven other predicted protein-coding genes and a putative ncRNA gene (*CR43491*) that likely corresponds to the *T1/T2* or *BarA* transcript identified previously (Higashijima, Kojima, *et al.* 1992; Norris *et al.* 1992). As predicted by Higashijima and colleagues (1992), the $B^1$ breakpoint lies in an intergenic region upstream of *B-H1* and downstream of *BarH2 (B-H2)* **(Figure 3B)**, a related homeodomain gene that is also involved in eye morphogenesis (Higashijima, Kojima, *et al.* 1992). Thus, the $B^1$ duplication on *FM7* chromosomes carries an intact *B-H2–B-H1* locus, plus an additional copy of *B-H1* fused downstream of *CG12432* **(Figure 3B)**.

Tsubota *et al.* (1989) proposed that unequal exchange between two *roo* insertions at different positions on homologous chromosomes caused the $B^1$ duplication **(Figure 3A)**. To provide an independent assessment of this hypothesis, we extracted split/discordant reads and their mate-pairs in the ±1.5-kb intervals at either end of the duplicated segment, then performed *de novo* assembly as above for the *FM7* inversions and recovered two contigs spanning the 16A1 and 16A7 sides of the $B^1$ breakpoint. Both of these contigs contained *roo* sequences that began after the exact point at which alignment to the reference genome ended. We used long-range PCR to amplify an approximately 8-kb fragment spanning the breakpoint from the end of the 16A7 to the beginning of 16A1 in *FM7*-carrying but not in wild type stocks **(Supplemental Table 2)**. Sanger sequencing of the 5' and 3' ends of this breakpoint-spanning fragment revealed a *roo* element in the expected location and orientation. Together, these results confirm the work of Tsubota *et al.* (1989), showing that the $B^1$ breakpoint contains a *roo* element in the 5' to 3' orientation located precisely at the junction between the duplicated segments.

Our genomic data also allows us to investigate sequence variation directly within the $B^1$ duplication, which provides new insights into the origin and reversion of the $B^1$ allele. Analysis of sequence variation in the region duplicated in $B^1$ revealed a large number of "heterozygous" SNPs in each hemizygous or homozygous *FM7* stock (min: 1242, max: 1250). "Heterozygous" SNPs in hemizygous or homozygous stocks can arise from calling variants in duplicated regions that are mapped to the same single-copy interval of the reference genome (Remnant *et al.* 2013). This apparent heterozygosity in the $B^1$ interval implies substantial sequence divergence existed between the two ancestral haplotypes that underwent unequal exchange to form the original $B^1$ allele, providing independent support for the origin of $B^1$ by unequal exchange between two homologous chromosomes rather than two sister chromatids (Tsubota *et al.* 1989). Additionally, the "heterozygous" SNP profile was nearly identical among all eight *FM7* stocks, supporting a single origin for the $B^1$ allele, consistent with the historical record (Tice 1914).

These "heterozygous" variants also give us a rich set of molecular markers that, together with depth of coverage in the *B* region, can be used to investigate the mechanism of $B^1$ reversion. If reversion of the $B^1$ allele is due to either inter-chromosomal or intra-chromosomal unequal exchange (Sturtevant and Morgan 1923; Sturtevant 1925; Peterson and Laughnan 1963; Gabay and Laughnan 1973), we expect a twofold reduction in the depth of coverage to be associated with loss of "heterozygosity" across the entire $B^1$ duplicated region in revertant chromosomes **(Figure 3C)**. To test this hypothesis, we identified two *X*-chromosome balancer stocks carrying reversions of $B^1$ (*Binsc-107-614* and *Binscy-107-624*) and sequenced their genomes. As expected, depth of coverage in both $B^1$ revertants was at wild type levels across the $B^1$ interval X:17,228,526–17,432,002 **(Figure 3B)**. Additionally, no high quality heterozygous SNPs or split/discordant reads were observed in the $B^1$ interval in either revertant.

Comparison of the single-copy haplotypes in the two revertants revealed likely sites of unequal exchange **(Figure 3D)**. *Binsc-107-614* and *Binscy-107-624* haplotypes in the $B^1$ interval contained the same SNPs from X:17,228,526–17,283,005 and again from X:17,388,394–17,432,002, but differed from each other in the central X:17,283,375–17,388,155 interval. This result indicated that unequal exchange must have occurred in a 370-bp window between X:17,283,005 and X:17,283,375 in one stock, and in a 239-bp window between positions X:17,388,155 and X:17,388,394 in the other stock. This result also implied that the haplotype from the beginning of $B^1$ to 17,283,005 is from the 5' duplicated segment, and the haplotype from X:17,388,394 to the end of $B^1$ is from the 3' duplicated segment. Because the SNPs defining the sites of unequal exchange were close to one another, we were able to phase haplotypes from the distal and proximal duplicates using read-pair data in non-recombinant *FM7* "heterozygotes". Knowing the phase and location of both non-recombinant haplotypes in the $B^1$ duplication allowed us to infer that unequal exchange occurred between X:17,283,005 and X:17,283,375 in *Binsc-107-614*, and independently between X:17,388,155 and X:17,388,394 in *Binscy-107-624*. Together, these data provide definitive genomic evidence that $B^1$ reversion is associated with unequal exchange among duplicated segments directly within the $B^1$ interval.

**DISCUSSION**

Our work provides detailed insight into the structure and diversity of the most commonly used *X*-chromosome balancer in *D. melanogaster, FM7*. We mapped and characterized breakpoints of the three large inversions present on *FM7* and identified major sequence differences in the vicinity of *g* between the two subtypes of *FM7* (*FM7a* and *FM7c*). Surprisingly, we identified megabase-scale tracts of unique sequence in different *FM7c*'s that likely arose from DCOs removing the female-sterile $sn^{X2}$ allele within the *In(1)dl-49* inversion. We further show that loss of the $sn^{X2}$ allele affects a substantial proportion of *FM7c* chromosomes at the Bloomington *Drosophila* Stock Center. Finally, we clarified the molecular organization of the $B^1$ allele carried on *FM7*, and provide definitive genomic evidence for origin and reversion of $B^1$ by unequal exchange. In contrast to the prevailing notion of balancers as clonal non-recombining chromosomes, our results provide evidence that rare recombination events have led to large-scale sequence differences among balancers currently used by *Drosophila* researchers.

Our work has a number of implications for the design and interpretation of experiments that use *X*-chromosome balancers in *D. melanogaster*. Knowing the precise molecular location of inversion breakpoints on *FM7* reveals regions of the *X*-chromosome that are more or less susceptible to exchange events. Furthermore, the fact that many *FM7c's* carry megabase-scale tracts of unique variation, and that a substantial proportion of *FM7* chromosomes are mislabeled, should motivate researchers to characterize which *FM7* subtype their stocks actually carry. The genomic scale of sequence differences between *FM7* subtypes is sufficiently large such that, without controlling properly for *FM7* subtype, effects attributed to balanced chromosomes in heterozygotes could arise from differences in the *FM7* genetic background. Our finding that reversion of the female sterile $sn^{X2}$ allele by DCO in the *In(1)dl-49* interval is common suggests researchers should be cautious when using *FM7c* for long-term stock maintenance of mutations in this region. We advise that replicate copies of such stocks be maintained and periodically checked for $sn^+$, $B^1$ males that could indicate breakdown of the balanced chromosome by a DCO event. Alternatively, such mutations could be maintained using attached-*X* stocks instead of balancer chromosomes (Ashburner *et al.* 2005). Unavoidable DCOs within the *In(1)dl-49* region that remove the $sn^{X2}$ allele on *FM7c* may motivate synthesis of a new generation of female-sterile *X*-chromosome balancers, perhaps by introducing additional inversions inside the *In(1)dl-49* interval on *FM7c*.

Our study also demonstrates the value of sequencing classical stocks of *D. melanogaster* to uncover the molecular basis of uncharacterized mutations and better understand the genetic background of mutant stocks. Despite the availability of a nearly complete, richly annotated genome sequence, over 1,000 existing classical mutations in *D.*

*melanogaster* have not been associated with gene models or linked to genomic sequences. Here we identified the causal molecular basis of three classical inversions (*In(1)sc$^8$*, *In(1)dl-49*, and *In(1)FM6*), mapped the locations of the *B$^1$* duplication and *Df(1)JA27* deletion, proposed candidates for the lesions that causes the *g$^4$* and *sn$^{X2}$* alleles, and identified an uncharacterized deletion in the *Df(1)JA27* chromosome. Further analysis of our genomic data should lead to insights about the molecular basis of additional mutations carried by these strains, including the sites of transgene insertions that mark some *FM7* balancer chromosomes (Casso *et al.* 2000; Le *et al.* 2006; Abreu-Blanco *et al.* 2011; Lattao *et al.* 2011; Pina and Pignoni 2012). Sequencing classical lab stocks can also lead to the identification of mislabelled strains (e.g. that *FM7a-23229* is in fact a *FM7c* chromosome) and unreported genotypes (e.g. *sn$^+$* in *FM7a-23229*), and thereby reduce sources of unwanted experimental variation. Systematic sequencing of stocks in the Bloomington *Drosophila* Stock Center could therefore improve the precision of *Drosophila* genetics and, in conjunction with extensive phenotypic information in FlyBase, provide a powerful model to develop workflows to identify rare disease variants in humans.

Future work on second and third chromosome balancers is needed to generalize the findings reported here, although such studies would be more challenging because genomic analysis would need to be performed in heterozygotes. Sequencing larger samples of *FM7* chromosomes could also provide deeper insight into the mechanisms of exchange in highly inverted chromosomes (Sturtevant and Beadle 1936; Novitski and Braver 1954). Because the 74 *FM7c sn*$^+$ stocks identified here are likely all *bona fide* *FM7c*'s that have undergone DCO with a balanced stock, these chromosomes should provide a rich sample to study how DCOs are distributed relative to the locations of breakpoints in inversion heterozygotes. Likewise, sequencing of additional $B^1$ revertants can now be used as a model system to study unequal exchange at the molecular level, especially given our finding that the two duplicated regions in $B^1$ differ by many variants. By generating a large sample of $B^1$ revertants in heterozygotes that differ from *FM7* outside the $B^1$ interval, it will be possible to precisely measure the relative contribution of inter- and intra-chromosomal unequal exchange events, and to understand how unequal exchange events are distributed across the duplicated region. More in-depth analysis of sequence variation among *FM7* chromosomes could also lead to insights about gene conversion between balancers and balanced chromosomes (Cooper *et al.* 2008; Blumenstiel *et al.* 2009), as well as whether the predicted accumulation of deleterious mutations on balancers is observed at the molecular level (Araye and Sawamura 2013). Finally, sequencing a larger panel of *FM7* chromosomes and more primitive *X*-chromosome balancers could shed light on the ancestral state of *FM7* at the time of its origin, and how inversions were integrated within inversions to create the founders of the *FM* series (Lewis and Mislove 1953).

**METHODS**

**Fly stocks used**

*X*-chromosome balancer stocks used in this experiment were obtained from either the Bloomington *Drosophila* Stock Center or from the *Drosophila* Genetic Resource Center (see **Supplemental Table 1** for stock identifiers). The $y^1$-$ncd^D$ stock that was used as a parental *X*-chromosome in the construction of the *ISO-1* reference genome strain (Brizuela *et al.* 1994) was obtained from Jim Kennison. Full genotypes of stocks are listed in **Supplemental Table 1** and are referred to in the text by their abbreviated genotype followed by their stock number (where available). All flies were kept on standard cornmeal-molasses and maintained at 25°C.

**DNA preparation and whole-genome sequencing**

DNA was prepared from 10 adult hemizygous *FM7*-carrying *Bar* eyed males for stocks *FM7a-785, FM7a-23229, FM7a-35522, FM7a-36489, FM7c-616, FM7c-3378, Binsc-107-614*, and *Binscy-107-624.* Because of the poor viability of *FM7*-carrying hemizygous males in *FM7c-5193* and *FM7c-36337*, DNA was prepared from a mixture of 10 adult hemizygous *FM7* male and homozygous *FM7* females for these two samples. Ten heterozygous adult females were used for the *FM7c-616*, *FM7c-5193*, *FM7c-36337*, and *FM7a-23229* heterozygous samples. Ten adult hemizygous yellow males were used for $y^1$-$ncd^D$ sample. All DNA samples were extracted using the Qiagen DNeasy Blood & Tissue Kit (catalog number 69504). Flies were starved for 4 hr before freezing at −80°C for at least 1 hr prior to DNA extraction. 600- to 800-bp fragments of DNA were selected after shearing and libraries were prepared using a Nextera DNA Sample Prep Kit (catalog number FC-121-1031) from Illumina following the manufacturer's directions. Hemizygous and homozygous males from stocks *FM7a-785, FM7a-23229, FM7a-35522, FM7a-36489, FM7c-616, FM7c-5193, FM7c-3378*, and *FM7c-36337* were sequenced as 100-bp paired-end samples on an Illumina HiSeq 2500. Heterozygous females from stocks *FM7c-616*, *FM7c-5193*, *FM7a-23229*, and *FM7c-36337* were sequenced as 150-bp paired-end samples on an Illumina HiSeq 2500. Hemizygous males from stocks $y^1$-$ncd^D$, *Binsc-107-614*, and *Binscy-107-624* were sequenced as 150-bp paired-end samples on an Illumina NextSeq.

**Genome alignment and SNP calling**

Alignment to the UCSC Genome Bioinformatics dm3 version of the Release 5 *D. melanogaster* reference genome sequence was performed using bwa (version 0.7.7-r441) (Li and Durbin 2009). Variants were called using SAMtools and BCFtools (version 0.1.19-44428cd) (Li, Handsaker, *et al.* 2009). Indels and low quality SNPs (qual<200) were filtered out of single-sample Variant Call Format (VCF) files. Unique SNPs were identified by additionally filtering out heterozygous SNPs from single sample VCF files and merging samples to identify SNPs present in only one sample using VCFtools (version 0.1.12b) and visualized as heatmaps using R (version 3.1.3).

**Identification, assembly, and validation of rearrangement breakpoints**

Rearrangement breakpoints were identified using three strategies. For the *In(1)sc$^8$*, *In(1)dl-49*, *In(1)FM6*, and *B* breakpoints, split/discordant *X*-chromosome read pairs were identified using samblaster (Faust and Hall 2014) and visualized using the UCSC genome browser (Rosenbloom *et al.* 2015). Clusters of split/discordant reads corresponding to putative breakpoints were identified in the approximate locations where rearrangements were expected based on classical work. Original fastq sequences of split/discordant reads and their mate-pairs from ±1.5kb around putative breakpoints from the same rearrangement were then merged from all eight *FM7* stocks into a single per-rearrangement file. SOAPdenovo2 (version 2.04) was then used to perform *de novo* assemblies for both breakpoints of each rearrangement at the same time using a kmer size of 41 or 51 for the *In(1)sc$^8$*, *In(1)dl-49*, and *In(1)FM6* inversions and a kmer size of 73 for the *B$^1$* duplication breakpoint (Li, Yu, *et al.* 2009). To identify the *In(1)dl-49* inversion, we also ran Breakdancer (version 1.4.4) (Chen *et al.* 2009) using default options with the exception that only the *X*-chromosome was analyzed (-o X) and any event with fewer than 10 supporting reads was ignored (-r 10). For the *B$^1$* duplication, we also identified an interval with the expected two-fold higher read-depth coverage in the location where the duplication was expected to be found (**Figure 3B**) (Lindsley and Zimm 1992).

Contigs spanning candidate breakpoint were used to design PCR primers on either side of each candidate breakpoint region using Primer3 (Rozen and Skaletsky 2000). PCR was performed using Phusion DNA polymerase (NEB, catalog #M0530L) using a 62$^o$C annealing temperature and 45 second extension time. PCR products were purified from a gel using a QIAquick PCR Purification Kit (Qiagen, catalog #28106) and Sanger sequenced. Long-range PCR of the junction between the two duplicated $B^1$ regions was performed using a Qiagen LongRange PCR Kit (catalog #206402) using 250 ng of genomic DNA, 59°C annealing temperature, and nine minute extension time.

**Screen for *sn* reversion in *FM7* stocks at the Bloomington *Drosophila* Stock Center.**

We visually screened 630 stocks from the Bloomington *Drosophila* Stock Center that were labeled as carrying *FM7c* for the presence or absence of the *sn* phenotype in *B* males. Eighty-two stocks yielded *B*, $sn^+$ males and were classified as putative *FM7c* revertants. To determine if putative *FM7c* revertants were in fact mislabeled *FM7a*'s, 79 of these putative *FM7c* revertants were screened for the presence of a diagnostic 24-bp deletion associated with the $g^4$ allele that is present on all *bona fide FM7c*'s. Primers used to amplify a fragment spanning the $g^4$ deletion were garnet_F2 (5'-ACACCCGCATCGTATTGATT-3') and garnet_R2 (5'-CCAGTTGGCTGAAACTGAAA-3'). DNA was prepared by placing single *B*, $sn^+$ males in a standard fly squish buffer (50 µL of 1M Tris pH 8.0, 0.5M EDTA, 5M NaCl) plus 1 µL of 10 mg/ml Proteinase K. Extracts were then placed in a thermocycler at 37°C for 30 minutes, 95°C for 2 minutes followed by a 4°C hold. PCR was performed using 4 µL of fly squish product in a total volume of 50 µL. Fragments were amplified using Phusion polymerase (NEB catalog number M0530L) reaction conditions were per manufactures instructions except for a 64°C annealing temperature, and 45 second extension time. PCR amplicons were Sanger sequenced and resulting sequences were aligned to the reference genome to determine the presence or absence of the 24-bp deletion.

**ACKNOWLEDGEMENTS**

We thank Jim Kennison for the $y^1$-$ncd^D$ stock; Kate Malanowski, Kendra Walton, and Anoja Perera for expert assistance with DNA sequencing; Angela Miller for assistance with editing and figure preparation; John Merriam, Dan Lindsley, Jim Kennison, Alexander Konev and Andreas Prokop for helpful discussions; members of the Hawley and Bergman laboratories for constructive comments on the manuscript; and Github for providing free private repositories that enabled this collaboration. DEM, NY, CBS, AJC and RSH were supported by the Stowers Institute for Medical Research; KRC was supported by NIH grant P40 OD018537. CMB was supported by Human Frontier Science Program Young Investigator grant RGY0093/2012.

**FIGURES**

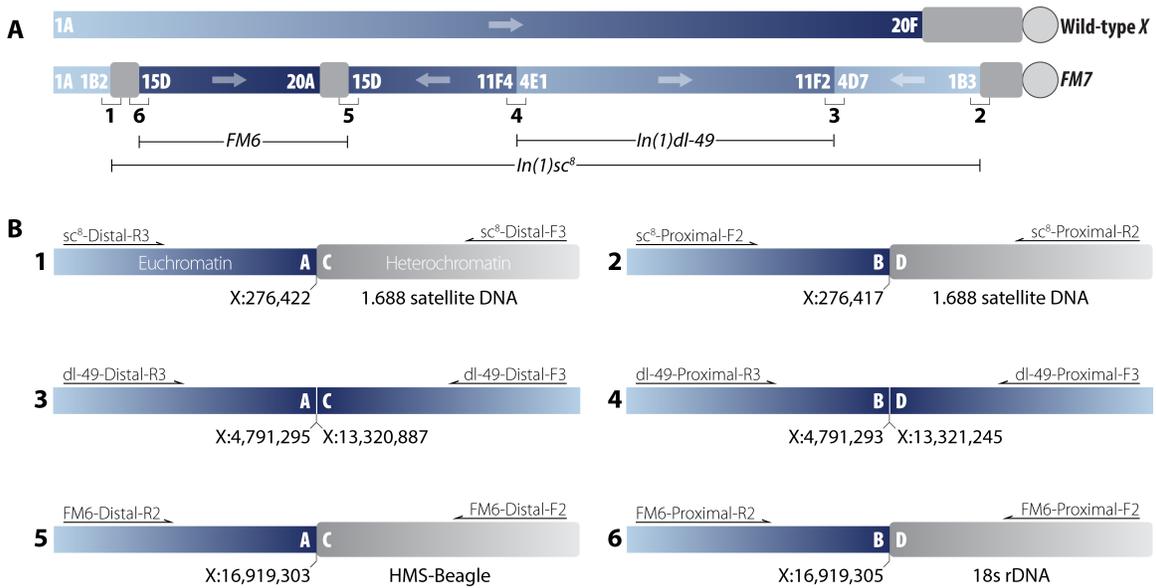

**Figure 1**. Structure of the *FM7* balancer chromosome. Euchromatin is shown in blue and heterochromatin is shown in grey. **A)** Schematic view of the organization of wild type and *FM7* X-chromosomes. *FM7* contains three inversions (*In(1)sc$^8$*, *In(1)dl-49*, and *In(1)FM6*) relative to wild type. The six breakpoint junctions for the three inversions are numbered 1-6 and are shown in detail in panel B. **B)** Location and organization of inversion breakpoints in *FM7*. Each inversion has two breakpoints that can be represented as A/B and C/D in the standard wild type arrangement and A/C and B/D in the inverted *FM7* arrangement, where A, B, C and D represent the sequences on either side of the breakpoints. Locations of euchromatic breakpoints are on Release 5 genome coordinates, and the identity of the best BLAST match in FlyBase is shown for heterochromatic sequences. Primers used for PCR amplification are shown above each breakpoint (see methods and **Supplemental File 2** for details). Forward and reverse primers are named with respect to the orientation of the assembled breakpoint contigs, not the orientation of the wild type or *FM7* X-chromosome.

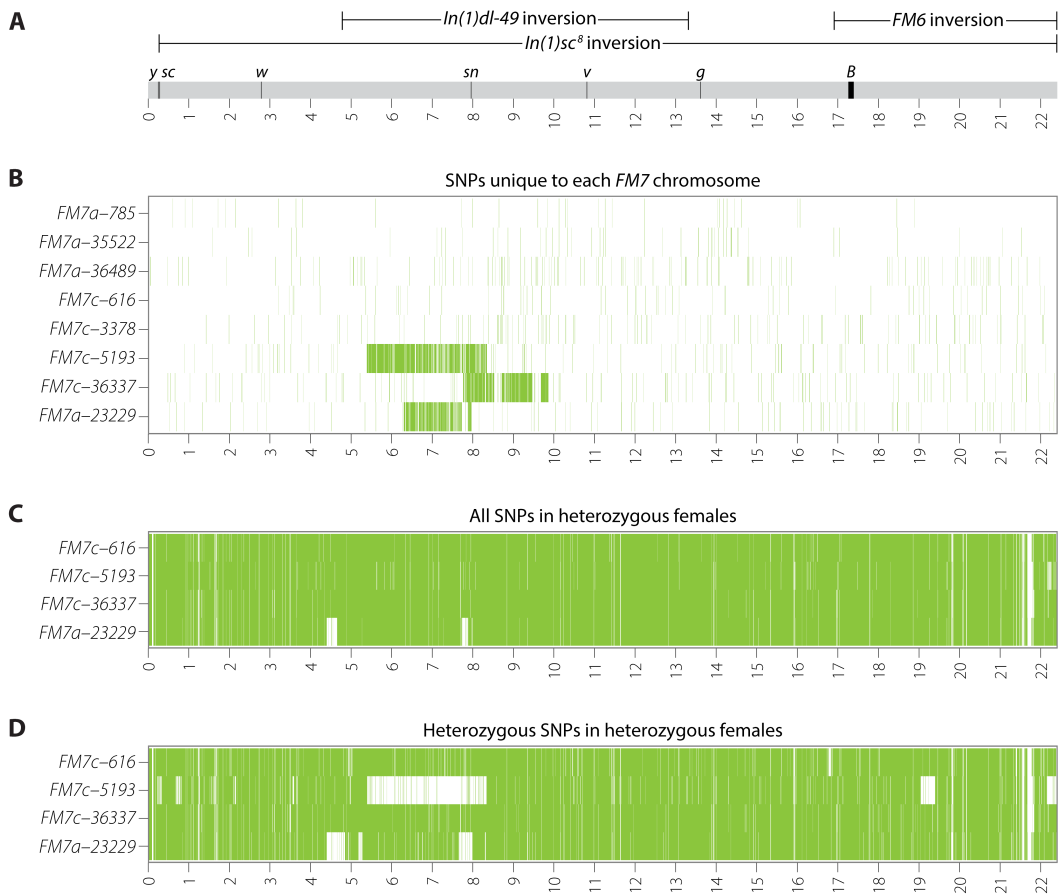

**Figure 2**. Recombination generates sequence diversity among *FM7* balancer chromosomes. **A)** Schematic of the wild type *X*-chromosome showing the locations of inversions (oriented with respect to the reference genome, not *FM7*), visible genetic markers, and Release 5 genome coordinates (in Mb). **B)** Heatmap of unique SNPs found in only one *FM7* chromosome in our sample. The density of unique SNPs is plotted in 5 kb windows with a 5 kb offset. The three large tracts of unique SNPs on *FM7c-5193*, *FM7c-36337*, and *FM7a-23229* all are contained fully within *In(1)dl-49* and replace the $sn^{X2}$ allele with wild type sequence. The *FM7a-23229* chromosome is a mislabeled *FM7c* (see **Supplemental Figure 1B**). **C)** Heatmap of all SNPs found in heterozygous female samples carrying *FM7* balancers over different balanced *X*-chromosomes. Genotypes of balanced *X*-chromosomes can be found in **Supplemental Table 1**. Small tracts where few SNPs are present in *FM7a-23229* arise because of common ancestry among the *X*-chromosomes in *FM7*, the balanced chromosome, and the *ISO-1* reference genome (see **Supplemental Figure 1C**). **D)** Heatmap of heterozygous SNPs found in heterozygous female samples carrying *FM7* balancers over different balanced X-chromosomes. Loss of heterozygosity (LOH) is observed for a large tract in *FM7c-5193* that corresponds to the large tract of unique variants for this chromosome shown in panel B. LOH is also observed in *FM7c-5193* for two deletions in the balanced chromosome (*Df(1)JA27* and an uncharacterized deletion on the *Df(1)JA27* chromosome), and for tracts in *FM7a-23229* that share ancestry with $y^1$-$ncd^D$ and *ISO-1* (see **Supplemental Figure 1C**).

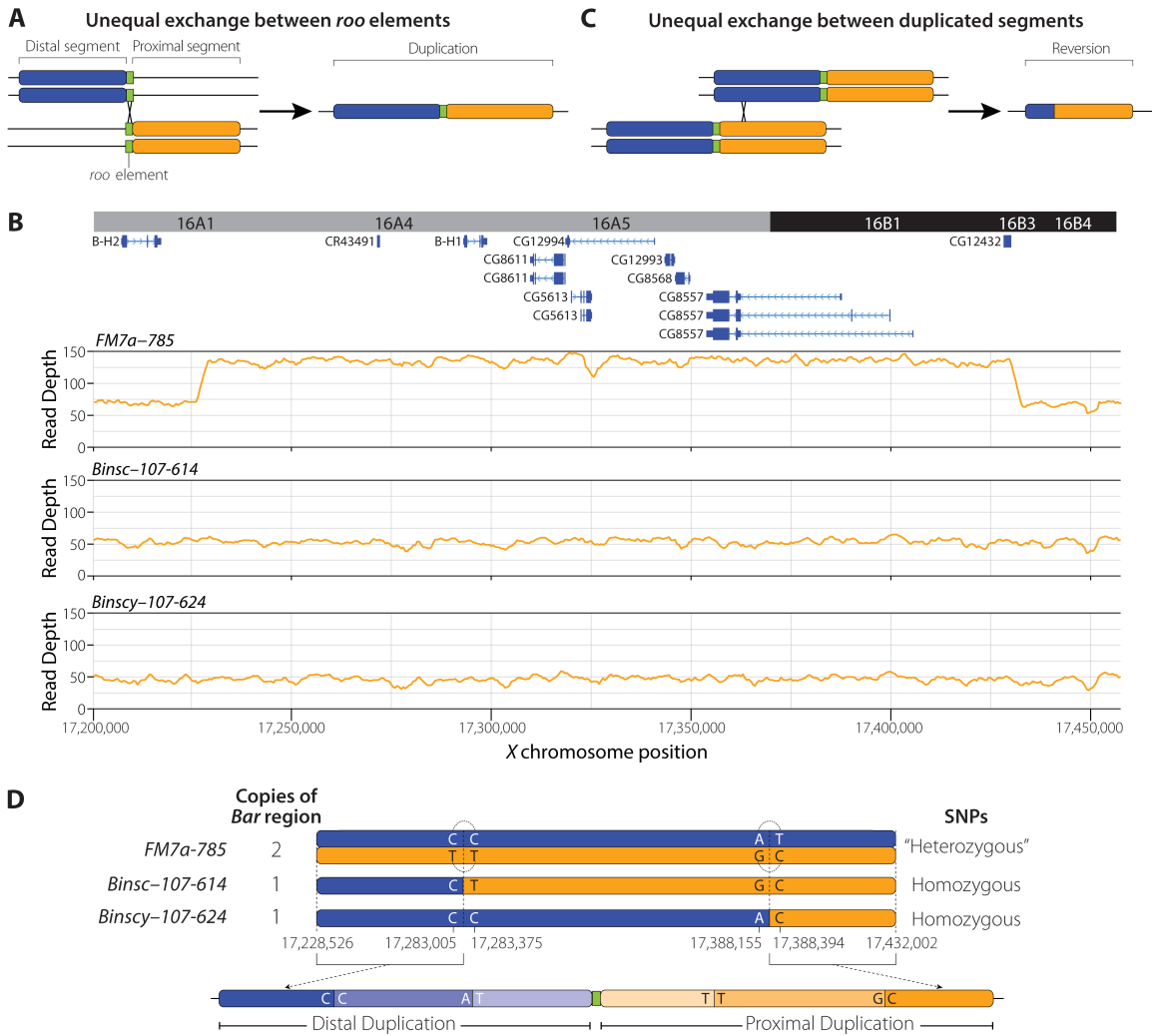

**Figure 3.** Genomic evidence for the role of unequal exchange at the *Bar* locus. **A)** Model for the origin of the $B^1$ allele by unequal exchange (Muller 1936) between two different *roo* transposable elements (Tsubota *et al.* 1989). The distal and proximal segments of the $B^1$ duplication are shown in blue and orange, respectively, and *roo* elements are shown in green. **B)** Genome annotation and depth of coverage for *X*-chromosome balancers carrying $B^1$ (*FM7a-785*) and wild type revertants (*Binsc-107-614* and *Binscy-107-624*). Note the twofold increase in depth that starts downstream of *B-H2* and ends upstream of *CG12432* in the *FM7a-785* chromosome carrying $B^1$ that is lacking in *Binsc-107-614* and *Binscy-107-624* revertants. **C)** Model for the reversion of the $B^1$ allele to wild type by unequal exchange between the two duplicated regions. The model shows an inter-chromosomal exchange event (Sturtevant and Morgan 1923; Sturtevant 1925) however intra-chromosomal exchange events are also possible (Peterson and Laughnan 1963; Gabay and Laughnan 1973). **D)** Schematic of sequence variants in $B^1$ chromosomes (*FM7a-785*) and wild type revertants (*Binsc-107-614* and *Binscy-107-624*). Sequences from the distal and proximal duplicated regions in $B^1$ chromosomes map to the same coordinates in the reference genome, resulting in apparent "heterozygosity". The two revertant chromosomes are characterized by different haplotypes of homozygous SNPs. Sequences shared by both revertants at their 5' and 3' ends can be used to define the boundaries of unequal exchange events and partially phase the distal and proximal haplotypes, respectively. Diagnostic SNPs from fragments that span the junctions of putative unequal exchange events can then be used to phase haplotypes on both sides of both exchange junctions in $B^1$ chromosomes (dotted arcs), which together with the sequence of the revertants, can be used assign the location of each exchange event to the appropriate revertant stock.

**SUPPORTING INFORMATION**

**Supplemental Figure 1**. Polymorphisms are evident both within *FM7* stocks and when comparing *FM7* stocks to the *ISO-1* reference genome. **A)** Schematic of the wild type *X*-chromosome, showing the locations of inversions (oriented with respect to the reference genome, not *FM7*), marker genes, and Release 5 genome coordinates (in Mb). **B)** Heatmap of SNPs detected in the eight *FM7* stocks used in this study when using *FM7a-785* as a genome reference. Increased SNP density covering the *sn* region in stocks *FM7c-5193*, *FM7c-36337*, and *FM7a-23229* indicates the region replaced by a DCO event. Note the increased SNP density between 13,369,185 Mb and 14,812,237 Mb present in all *FM7c* stocks (and the mislabeled *FM7a-23229*) that defines the haplotype containing $g^4$ present on *FM7c*. **C)** Heatmap of SNPs detected among all *FM7* stocks and $y^1$-*ncd$^D$* compared to the *ISO-1* reference genome. Sequence diversity among the eight *FM7* stocks is apparent at this scale as differing levels of SNP density surrounding the *sn* locus. Blocks of similarity between all *FM7*'s and *ISO-1* suggest a common ancestor for these regions. Blocks of diminished SNP density (in white) shared between *FM7a-23229* and $y^1$-*ncd$^D$* are apparent in Figure 2D as an apparent absence of SNPs.

**Supplemental Table 1**. Summary of genome sequencing for *D. melanogaster* stocks used in this study. Columns include data on stock names, stock center identifiers (where available), barcode used for multiplex sequencing, Stowers Institute LIMS and Flowcell identifiers, read length, average insert size, number of fastq file pairs, numbers of mapped reads overall and by chromosome arm, and average depth of coverage statistics by chromosome arm.

**Supplemental Table 2**. PCR primer and cycling conditions for *FM7* inversion and duplication breakpoints, and results of amplification in *FM7* and *ISO-1* wild type stocks.

**Supplemental Table 3**. Results of the *sn* reversion screen for *FM7* stocks in the Bloomington *Drosophila* Stock Center. Columns include data on the Bloomington *Drosophila* Stock Center stock identifier, the phenotype at the *sn* locus in *Bar*-eyed males ($sn^+$ or $sn^-$), and the presence or absence of a 24-bp deletion associated with the $g^4$ allele that marks *bona fide FM7c* chromosomes. Presence or absence of the $g^4$ allele at the molecular level was only determined for 79 of the 82 $sn^+$ stocks, and the $g^4$ status for rest of the stocks is not determined (n.d.).